 \documentclass[smallabstract,smallcaptions]{dccpaper}

\usepackage{epsfig}
\usepackage{amsmath}
\usepackage{amssymb}
\usepackage{color}
\usepackage{url}
\usepackage{svg}
\usepackage{booktabs}
\usepackage{multirow}
\usepackage{booktabs} \usepackage{makecell}
\usepackage{graphicx}
\usepackage{subcaption}
\usepackage[font=small,labelfont=bf]{caption} 

\newlength{\figurewidth}
\newlength{\smallfigurewidth}

\begin{document}

\title
{\large
\textbf{Partition Tree Search Acceleration for VVC: Survey and Evaluation with VTM Evolution}
}


\author{%
 M.E.A. Kherchouche$^{\ast \dag}$, F. Galpin$^{\ast}$, T. Dumas$^{\ast}$, L. Zhang$^{\dag}$, and D. Menard$^{\dag}$\\[0.5em]
{\small\begin{minipage}{\linewidth}\begin{center}
\begin{tabular}{ccc}
$^{\ast}$InterDigital, R\&I &&  $^{\dag}$Univ Rennes, INSA Rennes\\
845a Avenue des Champs Blancs && CNRS, IETR - UMR 6164 \\
35510 Cesson-Sevigne &&  F-35000 Rennes, France\\
\url{firstname.lastname@interdigital.com} && \url{firstname.lastname@insa-rennes.fr}
\end{tabular}
\end{center}\end{minipage}}
}

\maketitle
\thispagestyle{empty}

\begin{abstract}

The Versatile Video Coding (VVC) standard, introduced in 2020, offers 40–50\% of bitrate savings for equivalent visual quality of reconstructed videos over its predecessor High Efficiency Video Codec (HEVC), at the cost of significantly increased encoding complexity. This growth in encoding complexity is mainly due to the addition of the Quad Tree Multi Type Tree (QTMTT) partitioning structure which increase the split combinatorial.

This paper presents a critical evaluation of state-of-the-art (SOTA) partitioning acceleration techniques designed to reduce the complexity of the partitioning search in VVC. Particular attention is given to how these methods have evolved alongside successive versions of the VVC Test Model (VTM), which serves as the reference software for benchmarking coding tools. These techniques are analyzed in the context of their adaptation to internal changes in VTM, such as updated heuristics for fast partitioning decisions. The study also highlights the challenges involved in improving the trade-off between encoding complexity and compression efficiency. This challenge becomes more pronounced when evaluating methods across diverse VTM configurations and multiple software versions.

\end{abstract}

\section{Introduction}
Over the past decade, video content has exploded. According to recent forecasts, video now accounts for over 80\% of all the internet traffic worldwide, and this trend is only accelerating. At the same time, video characteristics have evolved towards complex formats, especially the move from Standard Definition (SD) to High Definition (HD) to Ultra High Definition (UHD), and the use of larger bit depths, e.g. \{8, 10\}-bit. This has multiplied the amount of data required per frame. Besides, the increase in frame rate, 120 Frames Per Second (FPS) and beyond, also augments the video data volume.
To address these challenges, the recently released Versatile Video Coding (VVC) \cite{VVC} standard dramatically improves coding efficiency over its predecessor High Efficiency Video Coding (HEVC) while supporting the needs for modern and future video applications. However, this coding efficiency comes with a jump in encoding complexity. This jump is caused by the expansion of candidate mode sets during the encoding process. 
Indeed, the encoding process is a combinatorial problem solved by the Rate-Distortion Optimization (RDO) process. The RDO selects the optimal encoding parameters by minimizing the Rate-Distortion (RD) cost across various encoding stages, including partitioning, intra/inter prediction, transform, quantization, and entropy coding. From HEVC to VVC, the expansion of candidate mode sets at every single stage has increased the RDO search space. 

Notably, at the partitioning stage which is the first RDO decision step, VVC introduces the Quad Tree Multi-Type Tree (QTMTT) structure, adding new split modes to those in HEVC. In addition to No Split (NS) and Quad Tree (QT), QTMTT includes Binary Tree (BT) and Ternary Tree (TT) splits. QT divides a coding unit (CU) into four equal sub-CUs; BT splits it into two equal parts horizontaly or verticaly (BTH/BTV); TT divides it into three sub-CUs using a $1\!:\!2\!:\!1$ size ratio horizontaly or verticaly (TTH/TTV).
Compared to HEVC, where a full RDO on a $64 \times 64$ intra Coding Tree Unit (CTU) involves testing up to $341$ blocks (i.e., reconstructing $20\cdot 10^3$ pixels per mode), VVC introduces much finer partitioning with QTMTT, resulting in up to $721\cdot10^3$ blocks and $19\cdot10^6$ pixels reconstructed per mode. This represents a worst-case increase of approximately $\sim2000\times$ more blocks and $\sim1000\times$ more pixels compared to HEVC.

In this paper, we propose a comprehensive survey of the VTM (VVC Test Model) evolution, along with a comparative analysis of partitioning acceleration techniques proposed by the SOTA. The study evaluates both the improvements across different VTM versions and the performance of the acceleration methods. In particular, a partitioning acceleration based on a size-independent Reinforcement Learning (RL)–based agent is evaluated through many VTM configurations and versions. 
This highlights VTM as a particularly challenging baseline for evaluating partitioning acceleration techniques, especially when multiple versions and carefully selected configurations are taken into account to ensure fair trade-offs between encoding complexity and compression efficiency.
The rest of the paper is organized as follows. Section \ref{Sec:VTM} introduces the VTM evolution and its various implementations. Section \ref{Sec:SOTAvsVTM} presents the SOTA partitioning acceleration techniques and compares them with the VTM evolution. Section \ref{Sec:RLvsVTM} focuses on the evaluation of the partitioning acceleration via a RL agent. Section \ref{Sec:Conclusion} concludes the paper.

\section{VVC Test Model (VTM) evolution}
\label{Sec:VTM}

Since the early stages of VVC development, the VTM has evolved alongside the standard, with continuous improvements in both algorithms and software structure, even after finalization. Early versions, such as VTM-3.0, included a limited set of coding tools and an incomplete RDO process, resulting in modest compression performance. These early implementations served primarily to validate future tools, rather than to reflect the efficiency or complexity of practical encoders.

\begin{table}[ht]
\centering
\captionsetup{font=scriptsize}
\caption{Encoding performance of different VTM versions relative to VTM-18.0 in AI configuration.}
\label{tab:VTMcpmlx}
\small
\resizebox{0.75\linewidth}{!}{%
\begin{tabular}{@{}lcccccccc@{}}
\toprule
\textbf{VTM Version}    & 3.0   & 5.0   & 6.0    & 7.0    & 8.0    & 9.0    & 10.2   & 23.11  \\ \midrule
\textbf{Relative CU Ratio (\%)} & 77.34 & 66.03 & 116.02 & 112.88 & 107.32 & 106.66 & 106.41 & 99.87 \\ 
\textbf{Relative Pixel Ratio (\%)} & 69.75 & 76.45 & 112.66 & 109.21 & 103.92 & 103.58 & 103.39 & 99.86 \\ 
\textbf{ET (\%)} & 75.80 & 65.50 & 102.20 & 102.70 & 105.00 & 103.20 & 103.10 & 93.50 \\ 
\textbf{BD-rate (Y) (\%)} & +8.60 & +4.14 & +1.72 & +1.57 & +1.97 & +0.67 & +0.65 & -0.05 \\
\textbf{BD-rate (YUV) (\%)} & +6.68 & +2.98 & +1.90 & +1.77 & +0.19 & +0.68 & +0.64 & -0.07 \\
\bottomrule
\end{tabular}
}
\end{table}

Table \ref{tab:VTMcpmlx} illustrates the evolution of several VTM versions considered in this survey, including VTMs: 3.0, 5.0, 6.0, 7.0, 8.0, 9.0, 10.2, 18 and 23.11. Table \ref{tab:VTMcpmlx} shows the CU-level complexity of each VTM version measured as the percentage of number of CUs evaluated during the RDO relative to the number evaluated in VTM-18.0 in the All Intra (AI) configuration. This complexity metric $C$ corresponds to the cumulated number \(C_x\) of reconstructed CUs through the RDO search at the encoder of a given VTM version against the cumulated number \(C_a\) of reconstructed CUs through the RDO search at the encoder of VTM-18.0.

{\small
\begin{equation}
\label{eq:CUsEvaluation}
    C = \frac{1}{4} \sum_{\text{QP}_i \in \{22, 27, 32, 37\}} \frac{C_a(\text{QP}_i) - C_x(\text{QP}_i)}{C_a(\text{QP}_i)}
\end{equation}
}

\(C_x(\text{QP}_i)\) and \(C_a(\text{QP}_i)\) refer to the values of \(C_x\) and \(C_a\) respectively when encoding at Quantization Parameter \(\text{QP}_i\). Another complexity metric uses the same formulation as in Eq.~\ref{eq:CUsEvaluation}, with the distinction that this complexity ratio compares the cumulated number of reconstructed pixels through the RDO search at the encoder of a VTM implementation against the cumulated number of reconstructed pixels through the RDO search at the encoder of this VTM with default maximum MTT depth 3.

Earlier versions, such as VTM-3.0 and VTM-5.0, incorporated fewer coding tools and simpler RDO, leading to lower CU evaluation rates but significantly higher mean BD-rate losses $+6.68\%$ and $+2.98\%$, respectively, compared to the reference version. Starting from VTM-6.0 and beyond, the integration of advanced tools increased the mode combinations, resulting in a noticeable increase in CU exploration, peaking at over 116\% in VTM-6.0 with a mean BD-rate loss of $+1.90\%$. 

\begin{figure}[ht]
    \centering
    \includegraphics[height=6cm, width=0.8\textwidth]{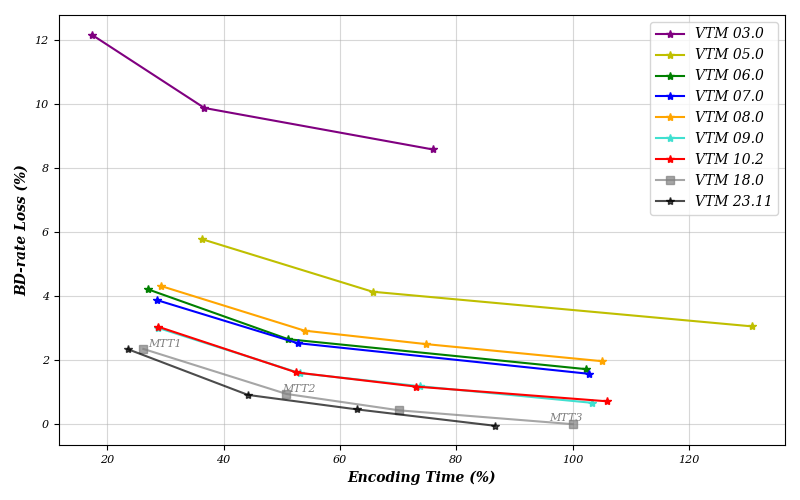}
    \captionsetup{font=scriptsize}
    \caption{Comparison of different VTM versions against VTM-18.0 with default encoding configuration in terms of encoding time and mean BD-rate.}
    \label{fig:VTMs}
\end{figure}

As the encoder matured, later versions (VTM~9.0 to VTM~23.11) incorporated more sophisticated heuristics and pruning mechanisms, progressively reducing unnecessary RDO evaluations while maintaining or even improving coding efficiency. Notably, VTM-8.0 introduced the Cross-Component Linear Model (CCLM) for chroma prediction, which enabled better exploitation of luma–chroma correlation. This enhancement allows a better UV reconstruction which then contributing to the sharp reduction in BD-rate for the YUV components in subsequent versions. 
VTM-23.11 is the latest implementation of VTM, which achieves quite a similar trade-off with VTM-18.0 in terms of CU evaluation complexity, while slightly outperforming it in overall BD-rate performance. This evolution of VTM highlights the progressive shift from a research-guided encoder to a more practical one, yet still a heuristic-driven implementation, which serves as the benchmark for evaluating emerging techniques such as machine learning-based decision frameworks.
Notably, the encoding time has decreased in newer versions like VTM-23.11, despite maintaining a similar number of RDO evaluations. This suggests that internal optimizations in the VTM implementation itself have contributed to reducing complexity.
In Fig. \ref{fig:VTMs}, the ratio $ET$ compares the encoding runtime \(T_x\) of the encoder of a given VTM version against the encoding runtime \(T_a\) of the encoder of VTM-18.0 with default encoding configuration.


{\small
\begin{equation}
    ET = \frac{1}{4} \sum_{\text{QP}_i \in \{22,27,32,37\}}{\frac{T_a(\text{QP}_i) - T_p(\text{QP}_i)}{T_a(\text{QP}_i)}}
\end{equation}
}

\(T_x(\text{QP}_i)\) and \(T_a(\text{QP}_i)\) refer to the values of \(T_x\) and \(T_a\) respectively when encoding at Quantization Parameter \(\text{QP}_i\).


Note that, in Fig. \ref{fig:VTMs}, for each VTM version, the points correspond to the points with the best trade-off between ET and mean BD-rate over many tests of encoding configurations. In each curve, three trade-off points, corresponding to three different values of the maximum MTT depths respectively are shown. The first point corresponds to the defaut VTM configuration, but modifying \texttt{MaxMTTHierarchyDepthI} = \texttt{MaxMTTHierarchyDepthISliceL} = \texttt{MaxMTTHierarchyDepthISliceC} = 1. The second point refers to defaut VTM configuration, but modifying \texttt{MaxMTTHierarchyDepthI} = \texttt{MaxMTTHierarchyDepthISliceL} = \texttt{MaxMTTHierarchyDepthISliceC} = 2. The third points is associated to default unchanged VTM configuration, i.e. \texttt{MaxMTTHierarchyDepthI} = \texttt{MaxMTTHierarchyDepthISliceL} = \texttt{MaxMTTHierarchyDepthISliceC} = 3. Since VTM-8.0, an intermediate point emerges, corresponding to default VTM configuration, but changing \texttt{MaxTTLumaISlice} = \texttt{MaxTTChromaISlice} = 16. All points are compared against the VTM-18.0 point associated to default encoding configuration, with $100\%$ encoding time and $0\%$ mean BD-rate loss.

\section{SOTA techniques against VTM evolution}
\label{Sec:SOTAvsVTM}

\begin{figure}[ht]
    \centering
    \includegraphics[height=8cm]{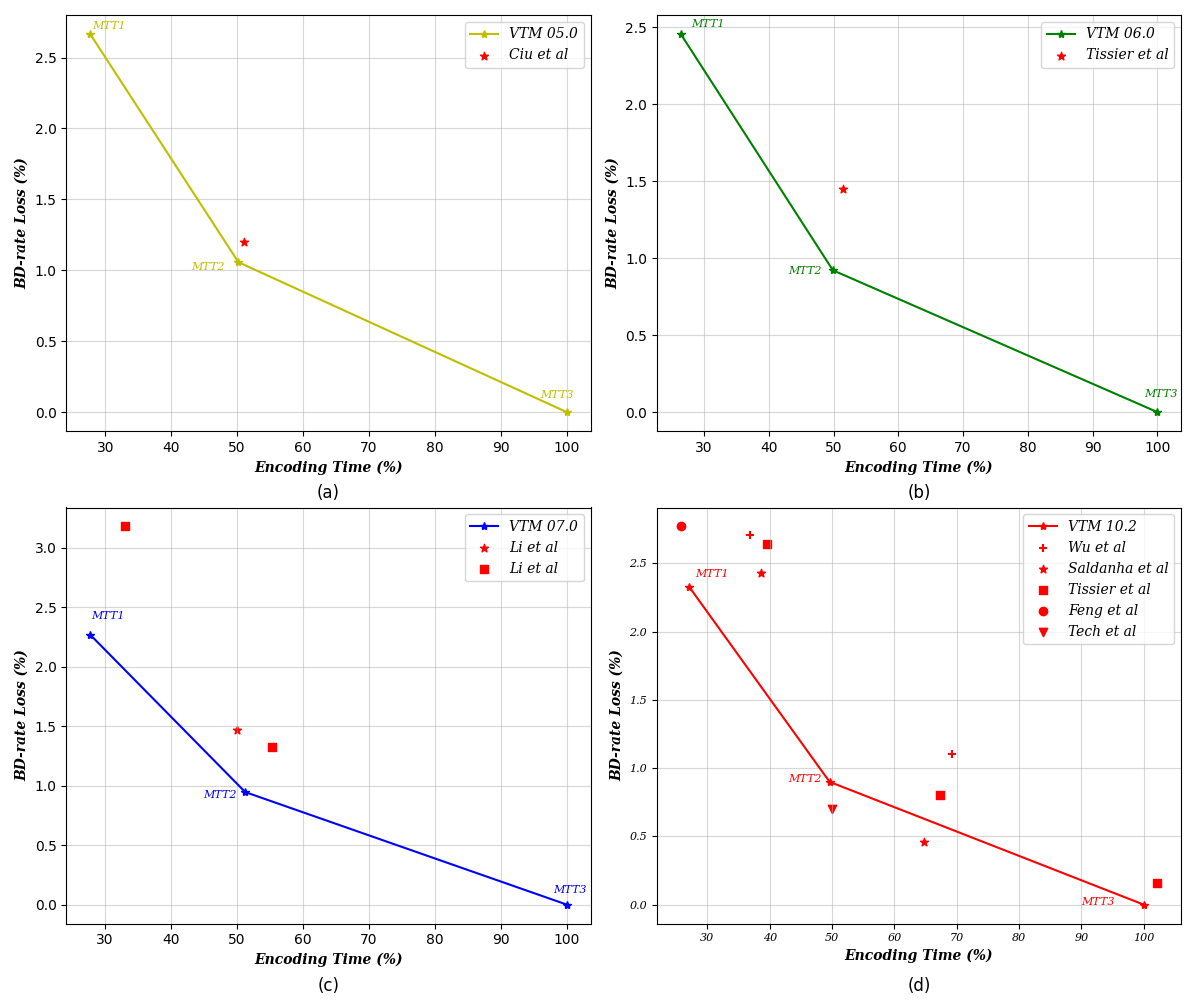}
    \captionsetup{font=scriptsize}
    \caption{SOTA trade-offs compared to different versions of VTM in AI configuration. \textbf{(a)} VTM-5.0 against the method proposed by  Cui \textit{et al.} \cite{VVC_HC5}. \textbf{(b)} VTM-6.0 against the method proposed by Tissier \textit{et al.} \cite{VVC3}. \textbf{(c)} VTM-7.0 against the methods \cite{HEVC7, VVC6}. \textbf{(d)} VTM-10.2 against methods proposed by \cite{VVC7, VVC8, VVC9, VVC_RDT}.}
    \label{fig:SOTAvsVTMs}
\end{figure}

Various acceleration techniques have been proposed across different video coding standards to reduce the exhaustive RDO searches. Partitioning methods have evolved significantly, leading to sophisticated and adaptive partitioning strategies. In VVC, numerous partitioning acceleration techniques exist, which have been implemented within the different VTM versions discussed above. 

Cui \textit{et al.} \cite{VVC_HC5} propose a gradient-based early termination method that uses directional texture analysis to predict partitioning decisions. By evaluating gradients in four directions, the algorithm selectively skips unlikely split types and reduces RDO evaluations. On top of VTM~5.0 in AI configuration, it achieves up to 51\% encoding time reduction with a 1.2\% mean BD-rate increase. Tissier \textit{et al.}~\cite{VVC3} propose a Convolutional Neural Network (CNN)-based approach to reduce partitioning complexity by predicting boundary probabilities in $64\times64$ CUs. These probabilities guide split decisions using adaptive thresholds based on CU size and split type. On top of VTM-6.1 in AI configuration, the method reduces encoding time by 51.5\% with a 1.45\% mean BD-rate increase. Li \textit{et al.} \cite{HEVC7} propose a pruning-based CNN framework that reduces model complexity by selectively removing weights layer by layer. Multiple pruned models are created to balance speed and RD performance, and a convex optimization scheme dynamically selects the best model per CTU. Tested on VTM-7.0 in AI configuration, the method halves encoding time with a 1.47\% mean BD-rate increase. The authors in \cite{VVC6}  propose a data-driven Mean Squared Error (MSE)-CNN that mimics QTMTT partitioning and employs early-exit to avoid redundant RDO. On top of VTM-7.0 in AI configuration, the method reduces encoding time by 44.65\% with a +1.32\% mean BD-rate increase. Saldanha \textit{et al.}~\cite{VVC7} propose a Light Gradient Boosting Machine (GBM)-based method that formulates block partitioning as five binary classification tasks, one per split type. Trained on features from texture and coding stats, the models guide split decisions in VTM-10.0, achieving 61.34\% encoding time reduction with 2.43\% mean BD-rate increase in AI configuration. Tissier \textit{et al.} \cite{VVC8} propose a two-stage approach combining a CNN for spatial feature extraction and a decision tree for MTT split prediction. On VTM-10.2 in AI configuration, it achieves up to 47.4\% encoding time reduction with 0.79\% mean BD-rate increase. Feng \textit{et al.}~\cite{VVC9} propose a CNN-based partition map prediction method to accelerate intra-frame encoding. On VTM~10.0 in AI configuration, the method reduces encoding time by 25.70\% with a 2.77\% mean BD-rate increase. Tech \textit{et al.} \cite{VVC_RDT}
propose a CNN-based method for fast intra-picture encoding, predicting partitioning parameters to minimize the rate-distortion-time cost. This method predicts the minimum width and height of sub-blocks in the MTT structure. The CNN is trained to output parameter pairs that control the partitioning of MTTs nested in QT nodes, enabling fine-grained control over the partitioning process. In VTM-10.0 in AI configuration, 50\% of average encoding runtime reduction with $+0.7\%$ of mean BD-rate is reported.
Fig.~\ref{fig:SOTAvsVTMs} presents the trade-offs achieved by the SOTA methods with respect to the VTM software version used in each case. The results are shown as individual points, each compared against the default VTM configuration adopted by the corresponding method (typically with a maximum MTT depth of 3). This highlights the importance of visualizing the performance of different VTM baselines under various VTM encoding configurations, at least different maximum MTT depths, to enable a fairer positioning of the SOTA methods. It is worth noting that all reported SOTA results are inherently tied to the specific VTM version and configuration used during their evaluation, which limits their relevance when compared across newer or differently configured versions of the reference software. 

Among the surveyed approaches, only the methods proposed by Tech \textit{et al.}  \cite{VVC_RDT} and Saldanha \textit{et al.} \cite{VVC7} consistently outperform the corresponding VTM baseline, underscoring their robustness across settings.

As stated in the updated JVET Call for Evidence (JVET-AM2026 \cite{JVET-AM2026}), evaluations are now carried out under multiple runtime conditions (from much faster to slower than VTM), which underlines the need to consider a range of trade-offs between encoding complexity and mean BD-rate rather than aiming at a single optimal point, unlike the above SOTA techniques.

\section{Example of partitioning acceleration technique through VTM configurations and versions}
\label{Sec:RLvsVTM}
In this section, a reinforcement learning (RL)–based partitioning acceleration method is evaluated across multiple VTM configurations and versions. RL is chosen for its flexibility to adapt to changing encoding settings and evolving software versions.

\subsection{RL agent for partitioning acceleration}
\vspace{-0.3cm}
\begin{figure}[ht]
    \centering
    \includegraphics[width=0.85\textwidth]{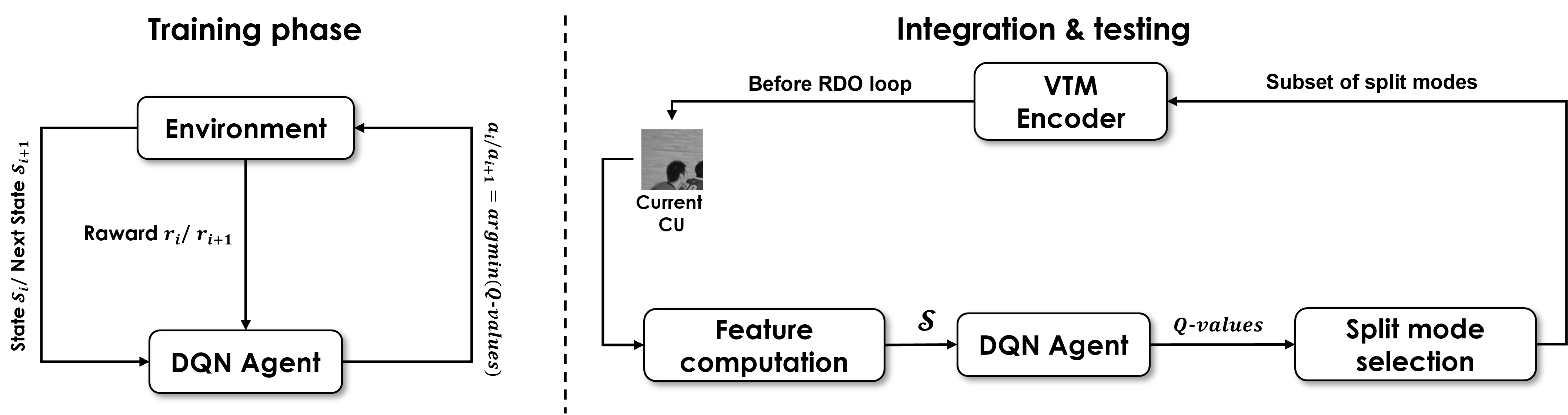}
    \captionsetup{font=scriptsize}
    \caption{Overview of the RL framework for CU partitioning. The RL agent receives a state vector extracted from the coding context and predicts Q-values that serve to select a subset of split modes, which are then applied to the VTM encoder.}
    \label{fig:RLagent}
\end{figure}

The block partitioning process in VVC is formulated as a Markov Decision Process (MDP), in which a Deep Q-Network (DQN) agent learns to approximate the Q-value function using the RD cost as a reward signal. For each possible split mode, i.e., NS, QT, BTH, BTV, TTH, TTV, a corresponding Q-value is predicted. The training is carried out using two-level partitioning trajectories, comprising both optimal and sub-optimal examples. The first level represents the root CU of size \( 32 \times 32 \), while the second level includes the sub-CUs resulting from each possible split.

As illustrated in Fig.\ref{fig:RLagent}, the proposed RL framework consists of separate training and inference phases. In both stages, each CU is represented as a fixed-dimensional feature vector \( \mathcal{S} \), which captures spatial, structural, and contextual information independently of the CU size. This unified representation enables the agent to generalize across multiple CU configurations that dominate partitioning complexity, namely within the set \( \mathcal{S}_{\text{size}} = \{32{\times}32, 16{\times}16, 32{\times}16, 16{\times}32, 8{\times}32, 32{\times}8\} \).


During training, the state vector \( \mathcal{S} \) is extracted from the environment using partitioning trajectories. In inference, it is generated by a dedicated feature computation module integrated into the VTM encoder. As shown in Table~\ref{tab:StateS}, \( \mathcal{S} \) combines four feature groups: Neighboring Information (NI), Parent Information (PI), Block Information (BI), and Spatial Information (SI). NI includes normalized RD costs and QT depths from top and left neighbors. To enrich spatial context, we incorporate their split series \( \text{SplitSeries}_{\text{Top}} \) and \( \text{SplitSeries}_{\text{Left}} \), representing the sequence of split modes from the CTU root to the neighbor. Each series is encoded as a 64-bit decimal, with 5-bit segments denoting partition types per depth. For each segment, a normalized pair \( (N_{\text{CU}}, HV) \) is derived, where \( N_{\text{CU}} \) is the number of sub-CUs (4 for QT, 2 for BT, 3 for TT), and \( HV \) indicates the orientation (0 for QT, +1 for horizontal, -1 for vertical).


The enriched representation enables the agent to capture both local RD behavior and hierarchical partitioning context. The resulting feature vector is fed into the DQN, which predicts Q-values guiding the partitioning decisions. During inference (see Fig.~\ref{fig:RLagent}), only the top \( N \in \{2,3,4,5\} \) split modes with the lowest predicted Q-values are evaluated by the RDO, reducing unnecessary exploration. A threshold \( T \in [0,1] \) can further restrict selection to candidates sufficiently close to the minimum Q-value.

\begin{table}[t]
\centering
\captionsetup{font=scriptsize}
\caption{Extracted Feature Sets for Fixed-Dimension Representation of a CU State.}
\label{tab:StateS}
\resizebox{0.7\textwidth}{!}{%
\begin{tabular}{cl}
\toprule
\textbf{Feature Group} & \textbf{Description} \\ \midrule
\textbf{NI}  & \makecell[l]{(1) Top RD cost, (2) Left RD cost, (3) Top QT depth, (4) Left QT depth,\\ (5) Top SplitSeries $(N_{CU}, HV)$, (6) Left SplitSeries $(N_{CU}, HV)$} \\ \midrule
\textbf{PI}  & \makecell[l]{(7) Parent NS RD cost, (8) NS Rate, (9) NS Distortion} \\ \midrule
\textbf{BI}  & \makecell[l]{(10) Width, (11) Height, (12) QP, (13) Current NS RD cost} \\ \midrule
\textbf{SI}  & (14) Histogram of Oriented Gradients (HOG) \\ 
\bottomrule
\end{tabular}%
}
\end{table}

The network is optimized using a composite loss function defined as:
\vspace{-0.3cm}
{\small
\begin{equation}
\mathcal{L} = \alpha_1 \cdot \text{MSE}_1 + \alpha_2 \cdot \text{MSE}_2 + \alpha_3 \cdot \text{MSE}_3
\end{equation}
}
The first term, \( \text{MSE}_1 \), supervises the Q-value predictions at the first partitioning depth, i.e. the root CU, by minimizing the error between predicted and ground-truth RD costs for all possible split actions.
\vspace{-0.4cm}
{\small
\begin{equation}
\text{MSE}_1 =  \left( \hat{\text{Cost}}_{l1}^{(i)} - \text{Cost}_{\text{GT},l1}^{(i)} \right)^2
\end{equation}
}
\( \hat{\text{Cost}}_{l1}^{(i)} \) and \( \text{Cost}_{\text{GT},l1}^{(i)} \) denote the predicted and true RD costs for action \( i \), respectively.
The second term, \( \text{MSE}_2 \), supervises the Q-values predicted at the second level of the tree for the sub-CUs, based on the action chosen at the root.
\vspace{-0.2cm}
{\small
\begin{equation}
\text{MSE}_2 =  \left( \hat{\text{Cost}}_{l2}^{(j)} - \text{Cost}_{\text{GT},l2}^{(j)} \right)^2
\end{equation}
}
\( \hat{\text{Cost}}_{l2}^{(j)} \) and \( \text{Cost}_{\text{GT},l2}^{(j)} \) are the predicted and true costs of the sub-CUs, respectively.

The third term, \( \text{MSE}_3 \), ensures hierarchical consistency between levels by aligning the Q-value of the parent CU with the reconstructed RD cost from its selected sub-CUs.
\vspace{-0.1cm}
{\small
\begin{equation}
\text{MSE}_3 = \left( \hat{\text{Cost}}_{\text{split}} - \left( \sum_{i} \hat{\text{Cost}}_{l2}^{(i)} + \Delta \text{Cost}_{\text{split}} \right) \right)^2
\end{equation}
}
Here, \( \hat{\text{Cost}}_{\text{split}} \) is the Q-value predicted at depth \( l1 \), \( \hat{\text{Cost}}_{l2}^{(i)} \) are the predicted Q-values for each sub-CU, and \( \Delta \text{Cost}_{\text{split}} \) is the actual syntax cost of signaling the split, preserved from the encoder.

This formulation is directly inspired by the Bellman equation:
{\small
\begin{equation}
Q(s, a) = r + \gamma \cdot \min_{a'} Q(s', a')
\end{equation}
}
where \( Q(s, a) \) is the expected cumulative cost of taking action \( a \) in state \( s \), \( r \) is the immediate RD cost, \( s' \) is the resulting next state, and \( \gamma \) is the discount factor (set to 1 in our hierarchical case). 

In this context, the temporal transition is replaced by a hierarchical transition across partition depths. Thus, \( \text{MSE}_3 \) serves as a Bellman-like Temporal Difference (TD) loss, adapted to a tree-structured decision process rather than a sequential one.
Training data is extracted from BVI-DVC~\cite{BVI-DVC} sequences encoded with VTM-18.0 in AI configuration and multiple maximum MTT depths, i.e., 4 and 6, resulting in over 12 million labeled $32{\times}32$ trajectories across the four QP values 22, 27, 32, and 37. An $\epsilon$-greedy strategy~\cite{Epsgreedy} is applied to balance exploration and exploitation.

\subsection{Experimental results}

This section evaluates the proposed RL agent by comparing its performance across three VTM versions: 10.2, 18.0, and 23.11. The agent is integrated into each version to ensure fair comparison. Experiments are conducted on the CTC test sequences~\cite{jvet:ctc}, grouped by resolution: A1/A2 ($3840\times2160$), B ($1920\times1080$), C ($832\times480$), and D ($416\times240$), and encoded at QPs 22, 27, 32, and 37. 


\begin{figure}[ht]
    \centering
    \begin{subfigure}[t]{0.48\textwidth}
        \centering
        \includegraphics[width=\linewidth]{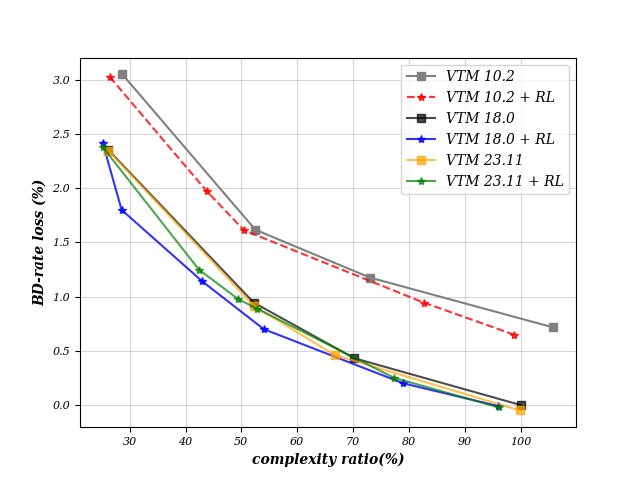}
        \caption{}
        \label{fig:RlVTM}
    \end{subfigure}
    \hfill
    \begin{subfigure}[t]{0.48\textwidth}
        \centering
        \includegraphics[width=\linewidth]{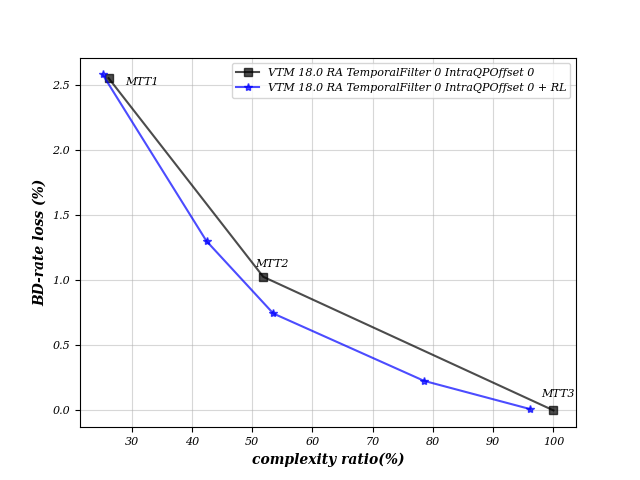}
        \caption{}
        \label{fig:RATest}
    \end{subfigure}
    \captionsetup{font=scriptsize}
    \caption{\textbf{(a)} Integration of the RL agent into the VTM versions 10.2, 18.0, and 23.11. \textbf{(b)} Integration of the RL agent into VTM-18.0 for the first intra frame in RA configuration.}
    \label{fig:RLIntegration}
\end{figure}

The complexity metric used for evaluation is based on the cumulative pixel ratio as discussed in Section \ref{Sec:VTM}. This formulation removes the dependency on the CU size and the VTM implementation and is based only on the RDO search.

\begin{table}[ht]
\centering
\captionsetup{font=scriptsize}
\caption{Trade-offs $N_{RL}$ in terms of mean BD-rate, average encoding runtime reduction $ET$, computational complexity reduction $Pixel_{ratio}$ and $CU_{ratio}$ against VTM-18.0 in AI configuration.}
\small
\resizebox{0.6\textwidth}{!}{
\begin{tabular}{ccccc}
\hline
 \textbf{``max. MTT depth''} & \textbf{BD-rate \%} & \textbf{$ET$ \%} & \textbf{$Pixel_{ratio}$\%} & \textbf{$CU_{ratio}$\%} \\ 
\hline

 111 & 2.42 & 27.60 & 25.16 & 15.59 \\ 
333 & 1.81  & 43.70 & 28.56 & 34.89 \\
 222 & 1.14 & 48.60 & 43.01 & 36.93 \\ 
 222 & 0.96 & 54.40 & 49.65 & 42.07  \\  
 333 & 0.71 & 67.60  & 54.07 & 58.96 \\ 
 333 & 0.21 & 89.90  & 79.06 & 80.47 \\
 333 & 0.01 &105.50  & 96.26  & 96.27\\
    
\hline
\end{tabular}
\label{tab:trade-offsQTMTT}
}
\end{table}
Fig.~\ref{fig:RLIntegration} presents the evaluation of the proposed RL agent integrated within multiple versions of the VTM reference software. In Fig.~\ref{fig:RLIntegration}(a), the trade-offs achieved by the agent are shown for VTM-10.2 (red), VTM-18.0 (blue), and VTM-23.11 (green), alongside their respective baselines (gray curves). All results are reported relative to the default configuration of VTM-18.0 at maximum MTT depth 3, serving as the reference point at 100\% complexity and 0\% BD-rate.
The results highlight the ability of the agent to provide a continuous and tunable trade-off curve across VTM versions, offering a level of adaptability not present in most SOTA methods, which typically yield a single fixed trade-off. However, while the RL agent introduces more flexibility, the best achievable trade-off points remain comparable to those reported by existing SOTA approaches. A summary of the RL trade-offs in terms of BD-rate, encoding time, and complexity is given in Table~\ref{tab:trade-offsQTMTT}.
Fig.~\ref{fig:RLIntegration}(b) further shows the robustness of the agent when applied to the Random Access (RA) configuration on VTM-18.0. Although trained only on the AI configuration, the agent generalizes well to RA first intra frames, producing trade-offs that outperform the RA baseline in a similar fashion. For instance, the agent achieves a $47\%$ complexity reduction ($Pixel_{ratio}$) with only $0.74\%$ BD-rate loss, and another point shows a $58\%$ reduction at $1.3\%$ BD-rate loss.

\section{Conclusion}
\label{Sec:Conclusion}

In this paper, a comprehensive evaluation of partitioning acceleration techniques in VVC was presented. The evolution of the VTM software was first reviewed, and its impact on encoding complexity across versions and configurations was analyzed. State-of-the-art speed-up methods were then surveyed and compared with respect to their compatibility with various VTM versions. Finally, a reinforcement learning–based approach was proposed, in which a size-independent DQN agent was designed to generalize across coding unit sizes and software versions. While the best trade-off points remain comparable to existing methods, greater flexibility and robustness were achieved across different encoding scenarios. These results indicate that learning-based strategies can be effectively integrated alongside traditional heuristic encoders to meet the evolving requirements of modern video coding standards.


\Section{References}
\bibliographystyle{IEEEtran}
\bibliography{refs}

@INPROCEEDINGS{VVC_HC5,
  author={Cui, Jing and Zhang, Tao and Gu, Chenchen and Zhang, Xinfeng and Ma, Siwei},
  booktitle={2020 Data Compression Conference (DCC)}, 
  title={Gradient-Based Early Termination of CU Partition in VVC Intra Coding}, 
  year={2020},
  volume={},
  number={},
  pages={103-112},
  doi={10.1109/DCC47342.2020.00018}}

@INPROCEEDINGS{VVC3,
  author={Tissier, A. and Hamidouche, W. and Vanne, J. and Galpin, F. and Menard, D.},
  booktitle={2020 IEEE International Conference on Image Processing (ICIP)}, 
  title={CNN Oriented Complexity Reduction Of VVC Intra Encoder}, 
  year={2020},
  volume={},
  number={},
  pages={3139-3143},
  doi={10.1109/ICIP40778.2020.9190797}}

@ARTICLE{HEVC7,
  author={Li, Tianyi and Xu, Mai and Deng, Xin and Shen, Liquan},
  journal={IEEE Transactions on Image Processing}, 
  title={Accelerate CTU Partition to Real Time for HEVC Encoding With Complexity Control}, 
  year={2020},
  volume={29},
  number={},
  pages={7482-7496},
  doi={10.1109/TIP.2020.3003730}}

@ARTICLE{VVC6,
  author={Li, Tianyi and Xu, Mai and Tang, Runzhi and Chen, Ying and Xing, Qunliang},
  journal={IEEE Transactions on Image Processing}, 
  title={DeepQTMT: A Deep Learning Approach for Fast QTMT-Based CU Partition of Intra-Mode VVC}, 
  year={2021},
  volume={30},
  number={},
  pages={5377-5390},
  doi={10.1109/TIP.2021.3083447}}

@ARTICLE{VVC7,
  author={Saldanha, Mário and Sanchez, Gustavo and Marcon, César and Agostini, Luciano},
  journal={IEEE Transactions on Circuits and Systems for Video Technology}, 
  title={Configurable Fast Block Partitioning for VVC Intra Coding Using Light Gradient Boosting Machine}, 
  year={2022},
  volume={32},
  number={6},
  pages={3947-3960},
  doi={10.1109/TCSVT.2021.3108671}}

@ARTICLE{VVC8,
  author={Tissier, Alexandre and Hamidouche, Wassim and Mdalsi, Souhaiel Belhadj Dit and Vanne, Jarno and Galpin, Franck and Menard, Daniel},
  journal={IEEE Transactions on Circuits and Systems for Video Technology}, 
  title={Machine Learning Based Efficient QT-MTT Partitioning Scheme for VVC Intra Encoders}, 
  year={2023},
  volume={33},
  number={8},
  pages={4279-4293},
  doi={10.1109/TCSVT.2022.3232385}}

@ARTICLE{VVC9,
  author={Feng, Aolin and Liu, Kang and Liu, Dong and Li, Li and Wu, Feng},
  journal={IEEE Transactions on Image Processing}, 
  title={Partition Map Prediction for Fast Block Partitioning in VVC Intra-Frame Coding}, 
  year={2023},
  volume={32},
  number={},
  pages={2237-2251},
  doi={10.1109/TIP.2023.3266165}}

@INPROCEEDINGS{VVC_RDT,
  author={Tech, Gerhard and Pfaff, Jonathan and Schwarz, Heiko and Helle, Philipp and Wieckowski, Adam and Marpe, Detlev and Wiegand, Thomas},
  booktitle={2021 Picture Coding Symposium (PCS)}, 
  title={Rate-Distortion-Time Cost Aware CNN Training for Fast VVC Intra-Picture Partitioning Decisions}, 
  year={2021},
  volume={},
  number={},
  pages={1-5},
  doi={10.1109/PCS50896.2021.9477452}}

@ARTICLE{VVC,
  author={Bross, Benjamin and Wang, Ye-Kui and Ye, Yan and Liu, Shan and Chen, Jianle and Sullivan, Gary J. and Ohm, Jens-Rainer},
  journal={IEEE Transactions on Circuits and Systems for Video Technology}, 
  title={Overview of the Versatile Video Coding (VVC) Standard and its Applications}, 
  year={2021},
  volume={31},
  number={10},
  pages={3736-3764},
  doi={10.1109/TCSVT.2021.3101953}}

@article{BVI-DVC,
	doi = {10.1109/tmm.2021.3108943},
  
	url = {https://doi.org/10.1109%2Ftmm.2021.3108943},
  
	year = 2022,
	publisher = {Institute of Electrical and Electronics Engineers ({IEEE})},
  
	volume = {24},
  
	pages = {3847--3858},
  
	author = {Di Ma and Fan Zhang and David R. Bull},
  
	title = {{BVI}-{DVC}: A Training Database for Deep Video Compression},
  
	journal = {{IEEE} Transactions on Multimedia}
}

@ARTICLE{jvet:ctc,
author = {Marta, Karczewicz and Yan, Ye},
  title = {Common test conditions and evaluation procedures for enhanced compression tool testing},
  journal={WG 05 MPEG Joint Video Coding Team(s) with ITU-T SG 16, $30^{th}$ meeting, Antalya},
    year = 	 "2023"
}

@book{Epsgreedy,
  title={Reinforcement Learning: An Introduction},
  author={Sutton, Richard S. and Barto, Andrew G.},
  year={2018},
  publisher={MIT Press},
  edition={2nd},
  url={https://web.stanford.edu/class/psych209/Readings/SuttonBartoIPRLBook2ndEd.pdf}
}

@techreport{JVET-AM2026,
  author       = {Jens-Rainer Ohm and Mathias Wien and Frank Bossen},
  title        = {Joint Call for Evidence on Video Compression with Capability Beyond VVC},
  institution  = {Joint Video Experts Team (JVET)},
  number       = {JVET-AM2026},
  year         = {2025},
  month        = {July},
  note         = {39th Meeting, Daejeon, KR, 26 June – 4 July 2025},
  url          = {https://jvet-experts.org/doc_end_user/documents/39_Daejeon/wg11/JVET-AM2026-v2.zip}
}

\end{document}